\documentclass[11pt,a4paper]{article} 
\usepackage{jheppub,hyperref,mdwlist}
\usepackage{amsmath}
\usepackage{graphicx}
\usepackage{subfigure}
\usepackage{amssymb}
\usepackage{xcolor}
\usepackage{multirow}
\usepackage{cancel}
\usepackage{color}
\usepackage{mathtools}
\usepackage{listings}
\usepackage{xcolor}
\usepackage{float}
\usepackage{slashed}

\usepackage{amsmath}
\usepackage{wrapfig}

\newcommand{\be}{\begin{equation}}
\newcommand{\ee}{\end{equation}}
\newcommand{\beq}{\begin{equation}}
\newcommand{\eeq}{\end{equation}}
\newcommand{\bea}{\begin{eqnarray}}
\newcommand{\eea}{\end{eqnarray}}
\newcommand{\besp}{\begin{equation}\begin{split}}
\newcommand{\eesp}{\end{split}\end{equation}}

\newcommand{\Br}{\text{Br}}

\newcommand{\Eq}[1]{Eq.~(\ref{#1})}

\newcommand{\Dfbd}{\mathord{\buildrel{\lower3pt\hbox{$\scriptscriptstyle\leftrightarrow$}}\over {D}_{\mu}}}

\def\mL{\mathcal{L}}
\def\mM{\mathcal{M}}

\def\mO{\mathcal{O}}

\def\mT{\mathcal{T}}

\def\Z{\mathbb{Z}}

\def\0{\textbf{0}}
\def\1{\textbf{1}}
\def\2{\textbf{2}}
\def\3{\textbf{3}}
\def\4{\textbf{4}}
\def\5{\textbf{5}}
\def\6{\textbf{6}}
\def\7{\textbf{7}}
\def\8{\textbf{8}}
\def\9{\textbf{9}}

\title{Probing electroweak phase transition with multi-TeV muon colliders and gravitational waves}

\author[a]{Wei Liu,}
\author[b]{Ke-Pan Xie}
\affiliation[a]{Department of Applied Physics, Nanjing University of Science and Technology, Nanjing 210094, People's Republic of China}
\affiliation[b]{Center for Theoretical Physics, Department of Physics and Astronomy, Seoul National University, Seoul 08826, Korea}

\emailAdd{wei.liu@njust.edu.cn}
\emailAdd{kpxie@snu.ac.kr}

\abstract{We study the complementarity of the proposed multi-TeV muon colliders and the near-future gravitational wave (GW) detectors to the first order electroweak phase transition (FOEWPT), taking the real scalar extended Standard Model as the representative model. A detailed collider simulation shows the FOEWPT parameter space can be greatly probed via the the vector boson fusion production of the singlet, and its subsequent decay to the di-Higgs or di-boson channels. Especially, almost all the parameter space yielding detectable GW signals can be probed by the muon colliders. Therefore, if we could detect stochastic GWs in the future, a muon collider could provide a hopeful crosscheck to identify their origin. On the other hand, there is considerable parameter space that escapes GW detections but is within the reach of the muon colliders. The precision measurements of Higgs couplings could also probe the FOEWPT parameter space efficiently.}

\begin{document}
\maketitle
\flushbottom

\section{Introduction}

Revealing the nature of the electroweak phase transition (EWPT) is one of the most important tasks in particle physics after the discovery of the Higgs boson at the LHC~\cite{Aad:2012tfa,Chatrchyan:2012ufa}. In the Standard Model (SM), lattice calculations have shown that the EWPT is a smooth crossover~\cite{Kajantie:1996qd,Rummukainen:1998as,Laine:1998jb}. However, the EWPT could be first-order (FO) in many new physics models beyond the SM (BSM), such as the real singlet extended SM (xSM)~\cite{McDonald:1993ey,Profumo:2007wc,Espinosa:2011ax,Cline:2012hg,Alanne:2014bra,Profumo:2014opa,Alves:2018jsw,Vaskonen:2016yiu,Huang:2018aja,Cheng:2018ajh,Alanne:2019bsm,Gould:2019qek,Carena:2019une,Ghorbani:2018yfr,Ghorbani:2017jls}, two-Higgs-doublet model~\cite{Turok:1990zg,Turok:1991uc,Cline:2011mm,Dorsch:2013wja,Chao:2015uoa,Basler:2016obg,Haarr:2016qzq,Dorsch:2017nza,Andersen:2017ika,Bernon:2017jgv,Wang:2018hnw,Wang:2019pet,Kainulainen:2019kyp,Su:2020pjw}, left-right symmetric model~\cite{Brdar:2019fur,Li:2020eun}\footnote{A research for the Pati-Salam model can be found in Ref.~\cite{Huang:2020bbe}.}, Georgi-Machacek model~\cite{Zhou:2018zli} and composite Higgs models~\cite{Espinosa:2011eu,Chala:2016ykx,Chala:2018opy,Bruggisser:2018mus,Bruggisser:2018mrt,Bian:2019kmg,DeCurtis:2019rxl,Xie:2020bkl}, etc. A FOEWPT can drive the early Universe out of thermal equilibrium, providing the essential environment for the electroweak baryogenesis (EWBG) mechanism~\cite{Morrissey:2012db,Cline:2006ts,Trodden:1998ym}, which explains the observed cosmological matter-antimatter asymmetry. 

The FOEWPT can manifest itself at two different kinds of experiments: the gravitational wave (GW) detectors and the high energy particle colliders. In the former case, the stochastic GWs generated during the FOEWPT are expected to be detectable at a few near-future space-based laser interferometers such as LISA~\cite{Audley:2017drz}, BBO~\cite{Crowder:2005nr}, TianQin~\cite{Luo:2015ght,Hu:2017yoc}, Taiji~\cite{Hu:2017mde,Guo:2018npi} and DECIGO~\cite{Kawamura:2011zz,Kawamura:2006up}. While in the latter case, the BSM physics related to FOEWPT might be probed at the colliders~\cite{Ramsey-Musolf:2019lsf}, such as the  CERN LHC and future proton-proton colliders including HE-LHC~\cite{Abada:2019ono}, SppC~\cite{CEPC-SPPCStudyGroup:2015csa} and FCC-hh~\cite{Benedikt:2018csr}, or future electron-positron colliders such as CEPC~\cite{CEPCStudyGroup:2018ghi}, ILC~\cite{Djouadi:2007ik} and FCC-ee~\cite{Abada:2019zxq}. Generally speaking, the hadron colliders have high energy reach but suffer from the huge QCD backgrounds, while the electron-positron colliders are very accurate but limited by the relatively low collision energy due to the large synchrotron radiation.

A muon collider might be able to offer both high collision energy and clean environment to probe the FOEWPT. On one hand, thanks to the suppressed synchrotron radiation compared to the electron, the energy of a muon collider can reach $\mO(10)$ TeV. What's more, the entire muon collision energy can be used to probe the short-distance reactions (hard processes). In contrast, at a $pp$ collider such as LHC, only a small fraction of the proton collision energy is available for the hard processes. On the other hand, due to the small QCD backgrounds, the muon collider is rather clean, allowing very precise measurements. The physics potential of a high energy muon collider has been discussed since the 1990s~\cite{Barger:1995hr,Barger:1996jm}, while it receives a renewed interest recently~\cite{Han:2012rb,Chakrabarty:2014pja,Ruhdorfer:2019utl,DiLuzio:2018jwd,Delahaye:2019omf,Long:2020wfp,Buttazzo:2018qqp,Costantini:2020stv,Han:2020uid,Capdevilla:2020qel,Han:2020uak,Han:2020pif,Bartosik:2020xwr,Chiesa:2020awd,Yin:2020afe,Buttazzo:2020eyl,Lu:2020dkx,Huang:2021nkl}.

In this work, we investigate the possibility of probing FOEWPT at a multi-TeV muon collider and the complementarity with the GW experiments, taking the xSM as the benchmark model. Although the xSM is simple, it has captured the most important features of the FOEWPT induced by tree level barrier via renormalizable operators~\cite{Chung:2012vg}, and can serve as the prototype of many BSM models that trigger the FOEWPT. For the muon collider setup, we follow Ref.~\cite{Han:2020pif} to consider collision energies of 3, 6, 10 and 30 TeV, with integrated luminosities of 1, 4, 10 and 90 ab$^{-1}$, respectively.

This paper is organized as follows. We first introduce the xSM and derive its parameter space for FOEWPT in Section~\ref{sec:xSM}, where we also discuss the GW signals and their detectability at the future LISA detector. The phenomenology at high energy muon colliders is studied in Section~\ref{sec:collider}, where both the direct (i.e. resonant production of the real singlet) and indirect (i.e. the Higgs coupling measurements) searches are considered. The complementarity between collider and GW experiments is also discussed. Finally, we conclude in Section~\ref{sec:conclusion}.

\section{FOEWPT in the xSM}\label{sec:xSM}

\subsection{The model}\label{sec:potential}

Up to renormalizable level, the scalar potential of xSM can be generally written as
\be\label{V}
V=-\mu^2|H|^2+\lambda|H|^4+\frac{a_1}{2}|H|^2S+\frac{a_2}{2}|H|^2S^2+b_1S+\frac{b_2}{2}S^2+\frac{b_3}{3}S^3+\frac{b_4}{4}S^4,
\ee
which has eight input parameters. However, one degree of freedom is unphysical due to the shift invariance of the potential under $S\to S+\sigma$; in addition, the measured Higgs mass $M_h=125.09$ GeV and vacuum expectation value (VEV) $v=246$ GeV put another two constraints, leaving us only five free physical input parameters.

To remove the shift invariance, we fix $b_1=0$ in \Eq{V}. In unitary gauge, \Eq{V} can be expanded around the VEV, i.e.
\be
H=\frac{1}{\sqrt2}\begin{pmatrix}0\\ v+h\end{pmatrix},\quad S=v_s+s,
\ee
and then the mass term of the two neutral scalars reads
\be
V\supset\frac12\begin{pmatrix}h&s\end{pmatrix}\mM_s^2\begin{pmatrix}h\\ s\end{pmatrix};\quad \mM_s^2=\begin{pmatrix}\frac{\partial^2V}{\partial h^2}&\frac{\partial^2V}{\partial h\partial s}\\ \frac{\partial^2V}{\partial h\partial s}& \frac{\partial^2V}{\partial s^2}\end{pmatrix}.
\ee
Diagonalizing $\mM_s^2$ yields the mass eigenstates $h_1$, $h_2$ and the mixing angle $\theta$ between them, namely
\be
\label{eq:trans}
\begin{pmatrix}h\\ s\end{pmatrix}=U\begin{pmatrix}h_1\\ h_2\end{pmatrix},\quad U=\begin{pmatrix}\cos\theta&-\sin\theta\\ \sin\theta&\cos\theta\end{pmatrix},
\ee
such that the mass matrix becomes $U^\dagger \mM_s^2U={\rm diag}\left\{M_{h_1}^2,M_{h_2}^2\right\}$. Here we assume the lighter state $h_1$ is the SM Higgs-like boson.

The requirement that $(v,v_s)$ is an extremum of \Eq{V} yields two relations~\cite{Alves:2018jsw}
\be\label{e1}
\mu^2=\lambda v^2+\frac{v_s}{2}(a_1+a_2v_s),\quad b_2=-\frac{1}{4v_s}\left[v^2(a_1+2a_2v_s)+4v_s^2(b_3+b_4v_s)\right],
\ee
where the coefficients $\lambda$, $a_1$ and $a_2$ can be further expressed in terms of $M_{h_1}$, $M_{h_2}$ and $\theta$,
\be\label{e2}\begin{split}
\lambda=&~\frac{M_{h_1}^2c_\theta^2+M_{h_2}^2s_\theta^2}{2v^2},\\
a_1=&~\frac{4v_s}{v^2}\left[v_s^2\left(2b_4+\frac{b_3}{v_s}\right)-M_{h_1}^2s_\theta^2-M_{h_2}^2c_\theta^2\right],\\
a_2=&~\frac{1}{2v_s}\left[\frac{s_{2\theta}}{v}\left(M_{h_1}^2-M_{h_2}^2\right)-a_1\right],
\end{split}\ee
with $c_\theta$ and $s_\theta$ being short for $\cos\theta$ and $\sin\theta$, respectively. Fixing $M_{h_1}=M_h=125.09$ GeV and $v=246$ GeV, we can use the following five parameters
\be\label{input}
\left\{M_{h_2},\theta,v_s,b_3,b_4\right\},
\ee
as input, and derive other parameters such as $\mu^2$, $\lambda$ via \Eq{e1} and \Eq{e2}.

We use the strategy described in Appendix~\ref{app:potential} to obtain the parameter space that satisfies the SM constraints. The dataset is stored in form of a list of the five input parameters in \Eq{input}, and then used for the calculation of FOEWPT and GWs in the following subsection.

\subsection{FOEWPT and GWs}

The scalar potential $V$ in \Eq{V} receives thermal corrections at finite temperature, becoming
\be\label{VT}\begin{split}
V_T=&-\left(\mu^2-c_HT^2\right)|H|^2+\lambda|H|^4+\frac{a_1}{2}|H|^2S+\frac{a_2}{2}|H|^2S^2\\
&+\left(b_1+m_1T^2\right)S+\frac{b_2+c_ST^2}{2}S^2+\frac{b_3}{3}S^3+\frac{b_4}{4}S^4,
\end{split}\ee
where we only keep the gauge invariant $T^2$-order terms~\cite{Dolan:1973qd,Braaten:1989kk}, and
\be
c_H=\frac{3g^2+g'^2}{16}+\frac{y_t^2}{4}+\frac{\lambda}{2}+\frac{a_2}{24},\quad
c_S=\frac{a_2}{6}+\frac{b_4}{4},\quad
m_1=\frac{a_1+b_3}{12}.
\ee
In our convention, $b_1=0$, the tadpole term for $s$ only arises at finite temperature. This term is found to be suppressed in most of the parameter space~\cite{Profumo:2007wc,Profumo:2014opa}, however for completeness we also include it in numerical study. As we will see very soon, the tadpole has a non-negligible impact on the FOEWPT pattern.

Thermal corrections change the vacuum structure of the scalar potential. In suitable parameter space, there exists a critical temperature $T_c$ at which the potential $V_T$ in \Eq{VT} has two degenerate vacua, one with $h=0$ (EW-symmetric) and the other with $h\neq0$ (EW-broken). Initially, the Universe stays in the EW-symmetric vacuum. As the Universe expands and the temperature falls below $T_c$, the $h\neq0$ vacuum is energetically preferred and the Universe acquires a probability of decaying to it. The decay rate per unit volume is~\cite{Linde:1981zj}
\be
\Gamma(T)\sim T^4\left(\frac{S_3(T)}{2\pi T}\right)^{3/2}e^{-S_3(T)/T},
\ee
where $S_3(T)$ is the Euclidean action of the $O(3)$-symmetric bounce solution. FOEWPT occurs when the decay rate per Hubble volume reaches $\mO(1)$ and hence the EW-broken vacuum bubbles start to nucleate. This defines the nucleation temperature $T_n$, which satisfies $\Gamma(T_n)=H^4(T_n)$, with $H(T)$ being the Hubble constant at temperature $T$. For a radiation-dominated Universe and a phase transition at EW scale, $T_n$ can be solved by the approximate relation~\cite{Quiros:1999jp}
\be\label{FOPT}
S_3(T_n)/T_n\approx140,
\ee
which we take as the criterion for a FOEWPT.

\begin{figure}
\centering
\subfigure{
\includegraphics[scale=0.4]{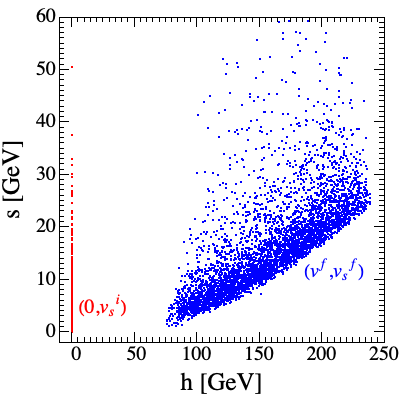}}\qquad\qquad
\subfigure{
\includegraphics[scale=0.4]{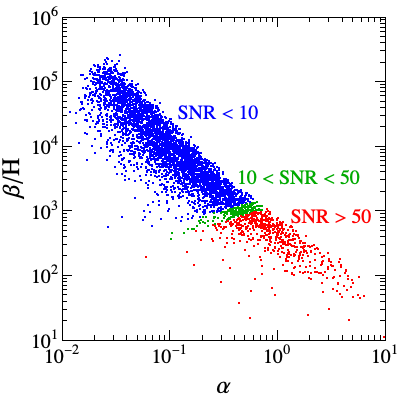}}
\caption{Left: the collection of vacuum decay initial states $(0,v_s^i)$ (red) and final states $(v^f,v_s^f)$ (blue). Right: the SNRs for the SFOEWPT data projected at the $\alpha$-$\beta/H$ plane.}
\label{fig:SFOEWPT}
\end{figure}

For each data point derived in last subsection, we calculate $T_n$ by solving \Eq{FOPT} with the {\tt Python} package {\tt CosmoTransitions}~\cite{Wainwright:2011kj}. Around $10\%$ of the data can trigger a FOEWPT. The left panel of Fig.~\ref{fig:SFOEWPT} shows the collection of FOEWPT data points by plotting the initial and final states of the vacuum decay $(0,v_s^i)\to(v^f,v_s^f)$. For a successful EWBG, the phase transition should be strong~\cite{Moore:1998swa,Zhou:2019uzq}
\be\label{strong}
v^f/T_n\gtrsim1,
\ee
such that the EW sphaleron process in the EW-broken vacuum is suppressed. Hereafter we will focus on data points satisfying \Eq{strong}. We found that many data points yield a decay pattern of $(0,0)\to(v^f,v_s^f)$, but there are also considerable fraction of data that have $v_s^i\neq0$. Quantitively, if we use $|v_s^i/v_s^f|\lesssim0.01$ as the criterion of a $(0,0)\to(v^f,v_s^f)$ FOEWPT, then the fraction of data points falling in this pattern is around $8.8\%$, while in Ref.~\cite{Alves:2018jsw} the corresponding fraction is $99\%$. We have checked that the difference comes from the treatment of the thermal tadpole term in \Eq{VT}: we keep this term, while Ref.~\cite{Alves:2018jsw} drops it. Therefore, the tadpole term actually has a considerable impact on the FOEWPT pattern.

A FOEWPT generates stochastic GWs mainly through three sources: bubble collisions, sound waves in the plasma and the magneto-hydrodynamics turbulence~\cite{Mazumdar:2018dfl}. After cosmological redshift, those GWs today typically peak at $f\sim{\rm mHz}$~\cite{Grojean:2006bp}, which is the sensitive region of a few next-generation space-based interferometers mentioned in the introuction. To obtain the GW spectrum today, we derive the following two parameters for each FOEWPT data point
\be
\alpha=\frac{1}{g_*\pi^2T_n^4/30}\left(T\frac{\partial\Delta V_T}{\partial T}-\Delta V_T\right)\Big|_{T_n};\quad \beta/H=T_n\frac{d(S_{3}/T)}{dT}\Big|_{T_n},
\ee
where $\Delta V_T=V_T|_{T_n,(v^f,v_s^f)}-V_T|_{T_n,(0,v_s^i)}$ is effective potential difference between the true and false vacua, and $g_*\sim100$ is the number of relativistic degrees of freedom. In other words, $\alpha$ is the transition latent heat over the radiation energy, while $\beta/H$ is the Universe expansion time scale over the phase transition duration. Refs.~\cite{Grojean:2006bp,Caprini:2015zlo,Caprini:2019egz} point out that the GW spectrum $\Omega_{\rm GW}(f)$ of a FOEWPT can be expressed as numerical functions of $(\alpha, \beta/H,v_b)$, where $v_b$ is the bubble expansion velocity. Taking $v_b=0.6$ as a benchmark, we are now able to calculate $\Omega_{\rm GW}(f)$ for each FOEWPT data point.\footnote{Note that $v_b$ is the bubble velocity with respect to the plasma {\it at finite distance}, while the velocity relevant for the EWBG calculation is actually $v_w$, which is defined as the relative velocity to the plasma {\it just in front of the wall}. The relation between $v_b$ and $v_w$ can be solved using hydrodynamics~\cite{Espinosa:2010hh,No:2011fi}, and it is possible to have a high $v_b$ (good for GW signals) and low $v_w$ (good for EWBG) simultaneously~\cite{No:2011fi,Alves:2018oct,Alves:2018jsw,Alves:2019igs,Alves:2020bpi}.} The suppression factor coming from the short duration of the sound wave period has been taken into account~\cite{Ellis:2018mja,Guo:2020grp}.

The signal-to-noise ratio (SNR) characterizes the detectability of GWs signals at an interferometer. Taking the LISA detector as an example, we calculate the SNR as
\be
{\rm SNR}=\sqrt{\mT\int_{f_{\rm min}}^{f_{\rm max}} df\left(\frac{\Omega_{\rm GW}(f)}{\Omega_{\rm LISA}(f)}\right)^2},
\ee
where $\Omega_{\rm LISA}$ is the sensitivity curve of the LISA detector~\cite{Caprini:2015zlo}, and $\mT=9.46 \times 10^7$ s the data-taking duration (around four years)~\cite{Caprini:2019egz}. According to Ref.~\cite{Caprini:2015zlo}, we use ${\rm SNR}>10~(50)$ as the detection threshold for a six-link (four-link) configuration LISA. In the right panel of Fig.~\ref{fig:SFOEWPT} we plot the $(\alpha,\beta/H)$ distribution as well as the SNRs of our data points. As shown in the figure, the data with large $\alpha$ (which means larger energy released in the transition) and smaller $\beta/H$ (means longer duration of the transition) have larger SNRs.\footnote{There are some data points with $\alpha\gtrsim1$, implying a strong supercooling. In this case, it is suggested that it is the percolation temperature $T_p$ rather than nucleation temperature $T_n$ that should be used to calculate $\alpha$ and $\beta/H$~\cite{Megevand:2016lpr,Kobakhidze:2017mru,Ellis:2018mja,Ellis:2020awk,Wang:2020jrd}. Since most of our data lie in the $\alpha\lesssim1$ region, we adopt the approximation $T_n\approx T_p$, and leave a more detailed treatment for the future work. Also note that we are using the traditional approach to derive FOEWPT profiles and calculate the GWs. It is shown that the alternative dimensional reduction approach can reduce the theoretical uncertainties significantly~\cite{Croon:2020cgk}.}

\section{Phenomenology at high energy muon colliders}\label{sec:collider}

Besides the GWs, the FOEWPT parameter space of the xSM can lead to signals of a resonantly produced heavy scalar (direct search), and corrections to the SM Higgs couplings (indirect search) at the colliders. The corresponding phenomenology has been studied at the LHC and the proposed $pp$ or $e^+e^-$ colliders~\cite{Profumo:2014opa,Cao:2017oez,Alves:2018oct,Zhou:2020idp,Alves:2018jsw,Chen:2019ebq,Huang:2016cjm,Alves:2019igs,Kozaczuk:2019pet,Papaefstathiou:2020iag,Alves:2020bpi,No:2018fev,Xie:2020wzn}. Typically, the direct search is implemented at $pp$ colliders due to their high energy reach, while the indirect approach is preferred by the $e^+e^-$ colliders because of the high accuracy. In this section we will demonstrate that, with the clean background and sufficient collision energy, a multi-TeV muon collider is able to perform both the direct and indirect searches, exhibiting a great potential to test the FOEWPT.

\subsection{Production and decays of the heavy scalar}

\begin{figure}
\centering
\includegraphics[width=0.385\textwidth]{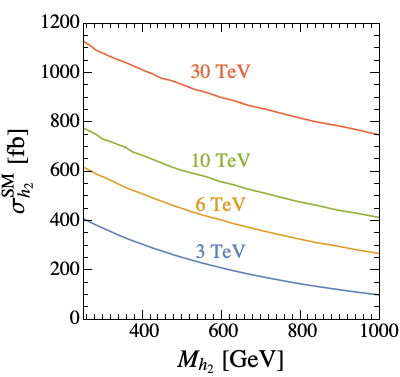}\qquad\qquad
\includegraphics[width=0.45\textwidth]{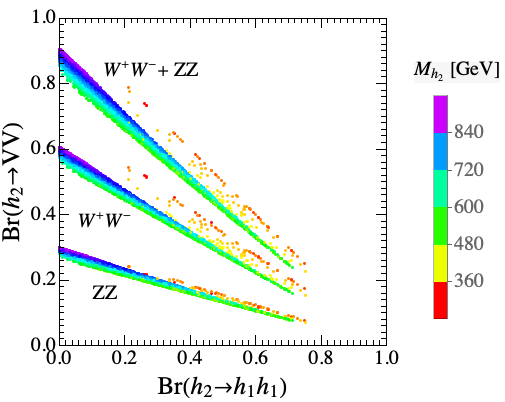}
\caption{Left: The SM-like production cross section of $h_2$ at muon colliders with different collision energies. Right: The scattering plots for the $\text{Br}(h_2 \rightarrow V V)$ and $\text{Br}(h_2 \rightarrow h_1 h_1)$ corresponding to the FOEWPT data points, and the value of $M_{h_2}$ is shown in color.}
\label{fig:xsec}
\end{figure}

The heavy scalar $h_2$ can be produced at a lepton collider via the $Zh_2$ associated production or the vector boson fusion (VBF) process. At a collider with center-of-mass energy as high as a few TeV, the dominant channel is VBF\footnote{The $\gamma h_2$ associated production (so-called ``radiative return'') can be comparable to the $Zh_2$ production at a low energy muon collider~\cite{Chakrabarty:2014pja}.}
\be\begin{split}
\mu^+ \mu^- \rightarrow&~ h_2 \nu_\mu \bar\nu_\nu\quad\text{($W^+W^-$ fusion)},\\ 
\mu^+ \mu^- \rightarrow&~h_2 \mu^{+} \mu^{-}\quad \text{($ZZ$ fusion)},
\end{split}\ee
and the production rate is
\be
\sigma_{h_2}=s_\theta^2\times\sigma^{\text{SM}}_{h_2},
\ee
where the SM-like production rate $\sigma^{\text{SM}}_{h_2} $ is the SM Higgs VBF production cross section evaluated by replacing the Higgs mass with $M_{h_2}$. This is because the coupling of $h_2$ to the SM gauge bosons comes from mixing, see \Eq{eq:trans}. In the left panel of Fig.~\ref{fig:xsec}, we have explicitly shown $\sigma^{\text{SM}}_{h_2}$ at the muon collider with different benchmark collision energies $\sqrt{s}$.\footnote{The cross sections are calculated by the {\tt MadGraph5aMC$@$NLO}-v2.8.2 event generator~\cite{Alwall:2014hca} with the model file written using {\tt FeynRules}~\cite{Alloul:2013bka}. The collider simulations in the next two subsections are also implemented with those two packages.} From the figure, it is clear that we can easily obtain hundreds of fb of cross section for $h_2$ with $\mO({\rm TeV})$ mass. For a given $\sqrt{s}$, the $W^+W^-$ fusion contributes $\sim90\%$ of the total cross section.

The produced $h_2$ will subsequently decays to multiple final states, such as di-Higgs ($h_1h_1$), di-boson ($W^+W^-$ and $ZZ$) and di-fermion (e.g. $t\bar t$). For the di-boson and di-fermion channels,
\be
\Gamma_{h_2 \rightarrow XX}=s_\theta^2 \times \Gamma^{\text{SM}}_{h_2 \rightarrow XX},
\ee
where $X$ denotes the SM vector boson or fermion, and $\Gamma^{\text{SM}}_{h_2 \rightarrow XX}$ is the decay width of the SM Higgs calculated at Higgs mass equal to $M_{h_2}$. For the di-Higgs channel,
\be\label{h2h1h1}
\Gamma_{h_2\to h_1 h_1}=\frac{\lambda_{h_2 h_1 h_1}^2 }{32 \pi M_{h_2}}\sqrt{1-\frac{4M^2_{h}}{M^2_{h_2}}},
\ee
where the $h_2h_1 h_1$ coupling is defined by
\be
\mL_{\rm xSM} \supset \frac{1}{2!}\lambda_{h_2 h_1 h_1}h_2h_1^2,
\ee
and at tree level~\cite{Huang:2016cjm}
\begin{multline}\label{lambdah2h1h1}
\lambda_{h_2 h_1 h_1}= \left(\frac{1}{2} a_1 + a_2 v_s\right) c_\theta^3 + (2 a_2 v-6 \lambda v) s_\theta c_\theta^2 \\
+\left(6 b_4 v_s + 2 b_3- 2 a_2 v_s - a_1\right) s_\theta^2 c_\theta- a_2 v s_\theta^3.
\end{multline}
The branching ratios are
\be\label{brh2h1h1}
\text{Br}(h_2 \rightarrow X X)=\frac{s_\theta^2 \times \Gamma^{\text{SM}}_{h_2 \rightarrow XX}}{s_\theta^2 \times \sum_{X'}\Gamma^{\text{SM}}_{h_2 \rightarrow X'X'} +\Gamma_{h_2\to h_1 h_1}},
\ee
\be\label{brh2h1h1}
\text{Br}(h_2 \rightarrow h_1 h_1)=\frac{\Gamma_{h_2 \rightarrow h_1 h_1}}{s_\theta^2 \times \sum_{X'}\Gamma^{\text{SM}}_{h_2 \rightarrow X'X'} +\Gamma_{h_2\to h_1 h_1} }.
\ee
The branching ratios of the FOEWPT data points are projected to the $\Br(h_2\to h_1h_1)$-$\Br(h_2\to VV)$ plane in the right panel of  Fig.~\ref{fig:xsec}, where $V=W^\pm$, $Z$. We see that the di-Higgs branching ratio can reach $\sim80\%$, while the $VV$ branching ratio dominates for large $M_{h_2}$. In general, all data points satisfy
\be
\Br(h_2\to h_1h_1)+\Br(h_2\to VV)+\Br(h_2\to t\bar t)\approx100\%,
\ee
and $\Br(h_2\to t\bar t)\lesssim20\%$. In the following two subsections, we choose the $h_2\to h_1h_1\to b\bar bb\bar b$ and $h_2\to ZZ\to\ell^+\ell^-\ell^+\ell^-$ as two complementary channels for collider simulations.

\subsection{Direct search: the $h_2\to h_1h_1\to b\bar bb\bar b$ channel}

Directly characterizing the portal coupling between the singlet and the Higgs boson, the $h_2 h_1 h_1$ coupling is of our primary interests. The signal of such a coupling is a resonant di-Higgs production at the muon collider, ${\rm VBF}\to h_2 \rightarrow h_1 h_1$. As the SM Higgs dominantly decays into $b \bar{b}$ pairs, the major final state of the signal consists of four $b$-jets which can be reconstructed into two $h_1$'s and then one $h_2$. For the $ZZ$ fusion production channel, the final state contains two additional forward muons. Here we focus on the so-called inclusive channel by including both the $W^+W^-$ fusion and $ZZ$ fusion events without detecting the additional muons. In this case, the main backgrounds are the SM VBF $h_1h_1$ and $ZZ$ production, with $h_1\to b\bar b$ and $Z\to b\bar b$.\footnote{There are also QCD backgrounds such as $4b2\nu_\mu$, but they turn out to be negligible after the Higgs candidates selection~\cite{No:2018fev,Han:2020pif}, thus will not be considered here.}

\begin{table}
	\small\centering\renewcommand\arraystretch{1.2}
	\begin{tabular}{c|ccccc}
		\hline
		Cross sections [ab] & $\sigma^{300}_S$ & $\sigma^{600}_S$ & $\sigma^{900}_S$ & $\sigma_{B}^{ZZ}$ & $\sigma_{B}^{h_1 h_1}$
		\\ \hline
		No Cut & 360 & 198 & 155 & 1080 & 567 \\
		Cut I  & 123 & 81.8  & 84.0 & 273 & 96.0\\
		Cut II  & 104 & 68.1 & 69.7 & 5.42 & 80.0 \\
		Cut III, 300  & 102 &   &  &  2.43 &  4.83 \\
		Cut III, 600  &  & 50.1 &   & $\mO(10^{-2})$ & 5.96\\
		Cut III, 900  &  &   & 35.7  & $\mO(10^{-2})$ & 2.96\\
		\hline
	\end{tabular}
	\caption{Cut flows at a 10 TeV muon collider for the signals with $M_{h_2}=$ 300, 600, 900 GeV and the backgrounds. For the signals, we have assumed $s_\theta =0.1$ and $\text{Br}(h_2 \rightarrow h_1 h_1) = 25 \%$.}
	\label{tab:SiD}
\end{table}

\begin{figure}
\centering
\includegraphics[width=0.445\textwidth]{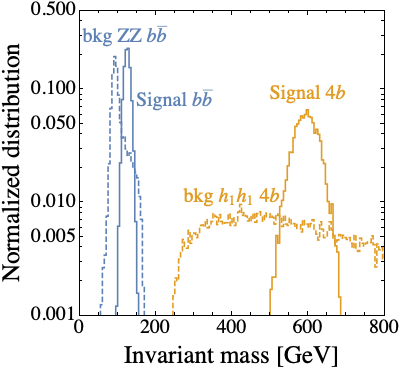}\qquad
\includegraphics[width=0.45\textwidth]{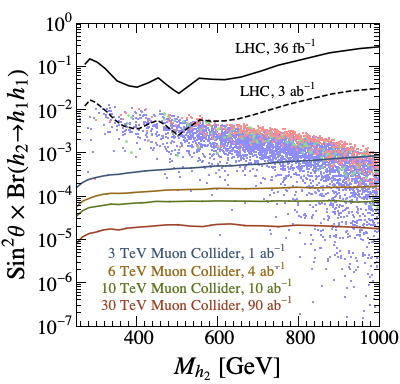}
\caption{Left: after the basic acceptance cuts, the invariant mass distributions of the jet pairs and four-jet system for the signal and main backgrounds at the 10 TeV muon collider. Here we select $M_{h_2} =$ 600 GeV as the signal benchmark. Right: the expected probe limits on $s_\theta^2 \times\Br(h_2 \rightarrow h_1 h_1)$ for different muon collider setups. The scatter points are the FOEWPT data, in which red, green and blue colors represent ${\rm SNR}\in[50,+\infty)$, $[10,50)$ and $[0,10)$, respectively. The limit from ATLAS at the 13 TeV LHC with $\mathcal{L}=$ 36.1 fb$^{-1}$~\cite{Aad:2019uzh} and its extrapolation to the HL-LHC~\cite{Alves:2018jsw} are also shown for comparison.}
\label{fig:hh}
\end{figure}

The signal and background events are generated at parton-level. We smear the jet four-momentum according to a jet energy resolution of $\Delta E/E=10\%$, and assume a conservative $b$-tagging efficiency rate of 70\%. The events are required to have exactly four $b$-jets satisfying the following basic acceptance cuts,
\be
p_T^j > 30~{\rm GeV},\quad  |\eta_j| < 2.43,\quad  M_{\text{recoil}}>200~\text{GeV}, \quad\text{(Cut I)}
\ee
where the pseudo-rapidity cut is based on a detector angular coverage of $10^\circ<\theta<170^\circ$, and the recoil mass is defined as
\be\label{M_recoil}
M_{\rm recoil}=\sqrt{\left(p_{\mu^+}+p_{\mu^-}-p_{j_1}-p_{j_2}-p_{j_3}-p_{j_4}\right)^2}.
\ee
Next, we pair the four $b$-jets by minimizing
\be
\chi_j^2=(M_{j_1 j_2}-M_{h})^2 + (M_{j_3 j_4}-M_{h})^2,
\ee
The pairs $(j_1,j_2)$ and $(j_3,j_4)$ are then identified as the Higgs candidates, in which the harder pair is defined as $(j_1,j_2)$. As shown in blue in the left panel of Fig.~\ref{fig:hh}, $M_{j_1j_2}$ and $M_{j_3j_4}$ peak around $M_{h}$ for the signal, while peak around $M_Z=91.188$ GeV for the $ZZ$ background. Therefore, a invariant mass cut
\be
|M_{j_1j_2}-M_{h} |<15~{\rm GeV},\quad |M_{j_3j_4}-M_{h} |<15~{\rm GeV}, \\
\quad\text{(Cut II)}
\ee
can significantly remove the $ZZ$ background. While most SM $h_1h_1$ events survive this cut, this background can be removed greatly by the cut on the four-jet system,
\be
|M_{4j}-M_{h_2}|<30~{\rm GeV},   \quad\text{(Cut III)}
\ee
as illustrated in orange in the left panel of Fig.~\ref{fig:hh}. The cut flows for three chosen signal benchmarks at a 10 TeV muon collider are shown in Table~\ref{tab:SiD}, indicating Cut III is fairly powerful to improve the signal over background factor.

Given the collision energy $\sqrt{s}$ and the integrated luminosity $\mL$, the signal and background event numbers are
\be\begin{split}
S=&~\sigma_S\times \epsilon_S\times \mL=\sigma^{\rm SM}_{h_2}\times s_\theta^2\times\Br(h_2\to h_1h_1)\times\epsilon_S\times\mL,\\
B=&~\sigma_B\times\epsilon_B\times\mL,
\end{split}\ee
where $\sigma_{S,B}$ are the signal and background production rates, and $\epsilon_{S,B}$ are the corresponding cut efficiencies, respectively. Note that $\sigma_B$ is already fixed, and $\sigma^{\rm SM}_{h_2}$ as well as $\epsilon_{S,B}$ depends only on $M_{h_2}$. This implies that we can generate events for several $M_{h_2}$ benchmarks and derive the collider probe limits for $s_\theta^2\times\Br(h_2\to h_1h_1)$, and make the interpolation to derive the $s_\theta^2\times\Br(h_2\to h_1h_1)$ reach as a function of $M_{h_2}$. As for the probe limits, we use the Poisson likelihood function
\begin{eqnarray}
L(S)= e^{(S+B)}	\frac{(S+B)^n}{n!}
\end{eqnarray}
with the number of observed events ($n$) taken to be equal to the background events ($n=B$). To get the 95\% confidence level exclusion limits, we use the test statistic $Q_{k}$
\begin{eqnarray}
Q_S \equiv -2 \ln\left[\frac{L(S)}{L(0)}\right]=3.84.
\end{eqnarray}
When $B\gg S$, the above procedure reduces to the well-known $S/\sqrt{B}=1.96$ criterion. The sensitivity of the muon collider to FOEWPT can be obtained by projecting the FOEWPT parameter space to such 2-dimension plane. This is done in the right panel of Fig.~\ref{fig:hh}, in which the reach of different collider setups are plotted as different colored solid lines, and the FOEWPT data points lying above a specific line can be probed by the corresponding muon collider. Note that our projections are derived with a rather conservative $b$-tagging efficiency of 70\%. A more optimistic efficiency such as 90\% can improve the results by a factor of 2, while an analysis without $b$-tagging will weaken the limits by a factor of 2 or 3, as in this case the non-$b$ jets (such as $W^\pm/Z\to jj$) also contribute to the backgrounds. However, either case has little visual effects in the log coordinate.

The right panel of Fig.~\ref{fig:hh} demonstrates that the FOEWPT parameter space can be greatly probed by the muon colliders, and higher energy colliders (with also higher integrated luminosities) give better reach. The current and projected LHC reach is shown in black lines for comparison. Because of the high accuracy in the multi-jet final state, even a 3 TeV muon collider (1 ab$^{-1}$) has a sensitivity more than one order of magnitude better than the HL-LHC (13 TeV, 3 ab$^{-1}$), and a 30 TeV muon collider (90 ab$^{-1}$) is able to probe $s_\theta^2\times\Br(h_2\to h_1h_1)$ up to $10^{-5}$, covering almost all of the FOEWPT parameter space. To manifest the complementarity with the GW experiments, we use different colors to mark the FOEWPT points with different SNRs: red, green and blue for ${\rm SNR}\in[50,+\infty)$, $[10,50)$ and $[0,10)$, respectively. Treating ${\rm SNR}=10$ as the detectable threshold, we see that those points which can be detected by LISA mostly lie in the reach of the muon colliders, especially for the $\sqrt{s}\geqslant6$ TeV setups. This is a great opportunity to identify the origin of the stochastic GWs, if they were detected in the future. On the other hand, the muon colliders also have significant sensitivity to the blue data points which are not detectable at the LISA. 

For muon colliders with $\sqrt{s}\leqslant10$ TeV, there are still appreciable number of points which can not be reached, due to the tiny $\text{Br}(h_2 \rightarrow h_1 h_1)$ in those points. Hence, in the next subsection, we will change our strategy by looking for a complementary decay channel, $h_2\to ZZ$, to finally cover those points. 

\subsection{Direct search: the $h_2\to ZZ\to\ell^+\ell^-\ell^+\ell^-$ channel}

The FOEWPT data points with tiny $\text{Br}(h_2 \rightarrow h_1 h_1)$ might be potentially probed via the $h_2 \rightarrow W^+W^-$ or $h_2\to ZZ$ channels. To get a better accuracy we focus on the leptonic decay of the gauge bosons. Although the $W^+W^-$ channel has a larger branching ratio, the neutrinos in the final state make this channel more challenging. In this subsection we would like to focus on the $ZZ$ channel with $Z\to\ell^+\ell^-$, where $\ell=e,\mu$, leading to a four-lepton final state.

\begin{figure}
\centering
\includegraphics[width=0.445\textwidth]{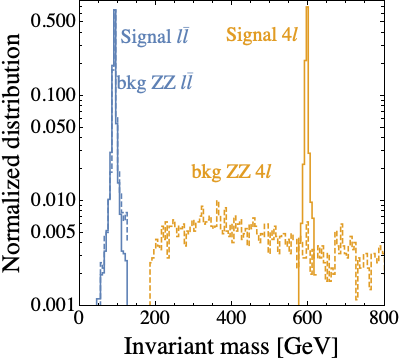}\qquad
\includegraphics[width=0.45\textwidth]{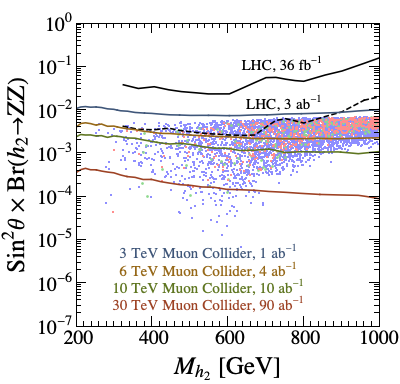}
\caption{Left: after the basic acceptance cuts, the invariant mass distributions of the lepton pairs and four-lepton system for the signal and main backgrounds at the 10 TeV muon collider. Here we select $M_{h_2} =$ 600 GeV as the signal benchmark. Right: the expected probe limits on $s_\theta^2 \times\Br(h_2 \rightarrow ZZ)$ for different muon collider setups. The scatter points are the FOEWPT data, in which red, green and blue colors represent ${\rm SNR}\in[50,+\infty)$, $[10,50)$ and $[0,10)$, respectively. The limit from ATLAS at the 13 TeV LHC with $\mathcal{L}=$ 36.1 fb$^{-1}$~\cite{Aaboud:2018bun} and its extrapolation to the HL-LHC~\cite{Alves:2018jsw} are also shown for comparison..}
\label{fig:ll}
\end{figure}

Similar to the treatment in the di-Higgs channel in the previous subsection, three cuts are applied to the events. We first require the events to have exactly four charged leptons with total zero charge, and satisfy the acceptance cuts
\be
p_T^\ell > 30~{\rm GeV},\quad  |\eta_\ell| < 2.43, \quad M_{\text{recoil}}>200~\text{GeV},\quad\text{(Cut I)}
\ee
where $M_{\rm recoil}$ is defined similarly to \Eq{M_recoil}. We then we pair the opposite-sign leptons by minimizing\footnote{We don't distinguish the lepton flavors here. We have checked that pairing method which distinguishes the flavors by classifying the same-flavor opposite-charge leptons into a pair [such as $(e^+e^-)(\mu^+\mu^-)$] gives almost the same $M_{\ell^+\ell^-}$ distributions.}
\be
\label{eq:4lpair}
\chi_\ell^2=(M_{\ell^+_1\ell^-_1}-M_Z)^2 + (M_{\ell^+_2\ell^-_2}-M_Z)^2,
\ee
and put a second cut
\be
|M_{\ell^+_1\ell^-_1}-M_Z |<10~{\rm GeV},\quad |M_{\ell^+_2\ell^-_2}-M_Z |<10~{\rm GeV},\quad\text{(Cut II)}
\ee
to the events. Note that this cut will select both the signal and background $ZZ$ events. Finally, a cut for the four-lepton system
\be
|M_{4\ell}-M_{h_2}|<20~{\rm GeV},\quad\text{(Cut III)}
\ee
can cut away the background significantly, as illustrated in the left panel of Fig.~\ref{fig:ll}.

The collider reach in $ZZ$ channel can be shown as $s_\theta^2\times \text{Br}(h_2 \rightarrow ZZ)$ limits in the right panel of Fig.~\ref{fig:ll}, in which different colored lines denote different muon collider setups, and the LHC current and future limits are also shown for comparison. Different from the case of $h_1 h_1$ channel, the reach of the HL-LHC is comparable to a 6 TeV muon collider (4 ab$^{-1}$) for $M_{h_2}\lesssim650$ GeV, thanks to the cleanness of the four-lepton final state. Nevertheless, better sensitivities can still be obtained for muon colliders with $\sqrt{s}\geqslant10$ TeV, especially for a 30 TeV (90 ab$^{-1}$) muon collider at which almost all the data points can be probed. We have checked that the FOEWPT data points escaping the $h_1h_1$ channel search can be generally reached in the $ZZ$ channel; as a result, the combination of $h_1h_1$ and $ZZ$ channels can cover almost the entire FOEWPT parameter space.\footnote{Fig.~\ref{fig:ll} shows that the 3 TeV muon collider (1 ab$^{-1}$) fails to probe any FOEWPT data points. This is because we only consider parameter space with a mixing angle $\theta\leqslant0.15$. For larger values of $\theta$ (such as 0.35 in Ref.~\cite{Alves:2018jsw}), there is reachable parameter space for the 3 TeV muon collider.} The complementarity with GW experiments are also shown in the figure.

\subsection{Indirect search: the Higgs coupling deviations}

\begin{figure}
\centering
\includegraphics[width=0.425\textwidth]{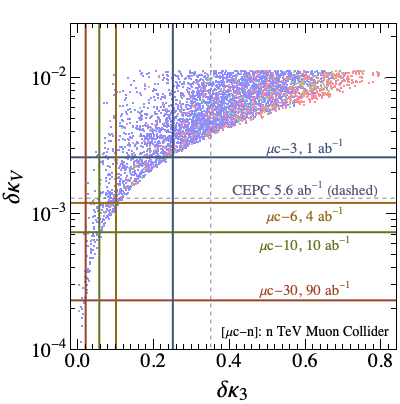}
\caption{Indirect limits from the measurements of the Higgs couplings. The scatter points are the FOEWPT data, in which red, green and blue colors represent ${\rm SNR}\in[50,+\infty)$, $[10,50)$ and $[0,10)$, respectively. The colored vertical and horizontal lines are the projections of different setups of muon colliders. The projections of CEPC ($\sqrt{s}=250$ GeV) are also shown in dashed lines for comparison.}
\label{fig:indirect}
\end{figure}

Besides the direct detection of the heavy scalar $h_2$, measuring the couplings of the Higgs-like boson $h_1$ also give hints of the FOEWPT, as those couplings usually deviate their corresponding SM values in the FOEWPT parameter space. For example, expanding the xSM Lagrangian as
\be
\mL_{\rm xSM}\supset\kappa_V\left(M_W^2W_\mu^+W^{-\mu}+\frac12M_Z^2Z_\mu Z^\mu\right)\frac{2h_1}{v}-\kappa_3\frac{M_h^2}{2v}h_1^3,
\ee
at tree level we obtain $\kappa_V=\kappa_3=1$ for the SM, while
\be
\kappa_V=c_\theta,\quad\kappa_3=\frac{2v}{M_h^2}\left[\lambda vc_\theta^3+\frac14c_\theta^2s_\theta\left(2a_2v_s+a_1\right)+\frac12a_2vc_\theta s_\theta^2+\frac13s_\theta^3\left(3b_4v_s+b_3\right)\right],
\ee
for the xSM. Defining the deviations as
\be
\delta\kappa_V=1-\kappa_V,\quad \delta\kappa_3=\kappa_3-1,
\ee
we project the FOEWPT data points into the $\delta\kappa_3$-$\delta\kappa_V$ plane in Fig.~\ref{fig:indirect}. One finds that $\delta\kappa_3$ is always positive (and $\lesssim0.8$). This can be understood by expanding the deviation at small mixing angle~\cite{Alves:2018jsw}
\be
\delta\kappa_3=\theta^2\left(-\frac32+\frac{2M_{h_2}^2-2b_3v_s-4b_4v_s^2}{M_h^2}\right)+\mO(\theta^3),
\ee
where the $M_{h_2}^2/M_h^2$ term dominates the terms in the bracket, implying an enhanced Higgs triple coupling. Since we set $\theta\leqslant0.15$ when scanning over the parameter space (see Appendix~\ref{app:potential}), the $\delta\kappa_V$ distribution has a sharp edge at around $0.15^2/2\approx0.01$.

Also shown in Fig.~\ref{fig:indirect} are the projections of the reach for different setups of muon colliders. The corresponding probe limits are adopted from Ref.~\cite{Han:2020pif}, which uses the VBF single Higgs production to study the $h_1VV$ coupling and the vector boson scattering di-Higgs production to study the triple Higgs coupling. It is clear that the FOEWPT parameter space can be probed very efficiently using via such indirect approach. A 3 TeV muon collider is already able to cover most of the data points, and a 30 TeV muon collider could test almost the whole parameter space.

\section{Conclusion}\label{sec:conclusion}

FOEWPT is an important BSM phenomenon that might exist in the early Universe, and shed light on today's collider and GW experiments. In this article, we perform a complementarity study of the proposed high energy muon colliders and the near-future space-based GW detectors to the FOEWPT. Choosing the xSM as the benchmark model, we first derive the FOEWPT parameter space, and then test the possibility of detecting it via GW signals or muon collider experiments. In the calculation of FOEWPT, we have included the thermal tadpole term, which is dropped in a few previous references. It is shown that the inclusion of tadpole term reduces the fraction of ``$(0,0)\to (v^f,v_s^f)$ pattern'' FOEWPT data points from 99\% to 8.8\%.

A considerable fraction of FOEWPT parameter space yields GW signals with ${\rm SNR}\geqslant10$ at the LISA detector, thus might be probed. Since the TianQin and Taiji projects in China have projected sensitivities similar to the LISA, we expect the FOEWPT parameter space could be probed and crosschecked by those detectors in the near future.

For the muon colliders, we consider center of mass energies $\sqrt{s}=3$, 6, 10 and 30 TeV, with corresponding integrated luminosities $\mL=1$, 4, 10 and 90 ab$^{-1}$, respectively. A detailed parton-level collider simulation is performed for the VBF production of the heavy real singlet and its subsequent decay to the di-Higgs ($h_2\to h_1h_1\to b\bar bb\bar b$) and di-boson ($h_2\to ZZ\to\ell^+\ell^-\ell^+\ell^-$) channels. The results in Figs.~\ref{fig:hh} and \ref{fig:ll} show that muon colliders offer great opportunity to probe the FOEWPT parameter space. In the di-Higgs channel, all muon collider setups have much higher reach compared to the HL-LHC; while in the di-boson channel, the reach of HL-LHC is comparable with a 6 TeV muon collider, however the muon colliders with $\sqrt{s}\geqslant10$ TeV still work much better. Combining the di-Higgs and di-boson channels allows us to probe the whole parameter space. In addition, given the high accuracy of the muon colliders, the precision measurements of the Higgs gauge and triple couplings also help to test the FOEWPT.

As for the complementarity with the GW experiments, a remarkable result is that almost all parameter space yielding detectable GWs is within the reach of the muon colliders. This implies if in the future we really detected some signals at the GW detector, a muon collider could provide very useful crosscheck to locate their origin. We also find that there is large parameter space that is not detectable via GWs, but can be probed at the muon colliders.

\acknowledgments

We are grateful to Huai-Ke Guo and Ligong Bian for the very useful discussions and sharing the codes. KPX is supported by the Grant Korea NRF-2019R1C1C1010050.

\appendix
\section{Deriving the phenomenologically allowed potential}\label{app:potential}

This appendix demonstrates how to derive the parameter space of a xSM scalar potential that satisfies current phenomenological bounds. In summary, our strategy contains two steps:
\begin{enumerate}
\item Construct a potential in \Eq{V}, and make sure it has a VEV in $(h,s)=(v,0)$, and a Higgs mass $M_{h_1}=M_h$. In this case, generally $b_1\neq0$.
\item Shift the $s$ field such that $b_1=0$, and match the new coefficients to the ones described in \Eq{input} of Section~\ref{sec:potential}. In this case, generally $v_s\neq0$.
\end{enumerate}
For the first step, an extremum at $(v,0)$ requires~\cite{Lewis:2017dme}
\be\label{a1}
\mu^2=\lambda v^2,\quad b_1=-\frac{v^2}{4}a_1;
\ee
and other coefficients can be expressed by the mass eigenvalues and mixing angle
\be\label{a2}
a_1=\frac{s_{2\theta}}{v}\left(M_{h_1}^2-M_{h_2}^2\right),\quad b_2+\frac{a_2}{2}v^2=M_{h_1}^2s_\theta^2+M_{h_2}^2c_\theta^2,\quad\lambda=\frac{1}{2v^2}\left(M_{h_1}^2c_\theta^2+M_{h_2}^2s_\theta^2\right).
\ee
Therefore, we can use
\be\label{app_input}
\left\{M_{h_2},\theta,a_2,b_3,b_4\right\},
\ee
as input and derive other coefficients via \Eq{a1} and \Eq{a2}. We randomly generate the input parameters in the following range:\footnote{If $h_2$ is too heavy, its thermal effect will be suppressed by the Boltzmann factor and the high-temperature expansion doesn't apply. Therefore we require $M_{h_2}\leqslant1$ TeV.}
\be\begin{split}
&M_{h_2}\in[250,1000]~{\rm GeV},\quad \theta\in[0,0.15],\\
&b_4\in[0,4\pi/3],\quad a_2\in[-2\sqrt{\lambda b_4},4\pi],\quad |b_3|\in[0,4\pi v],
\end{split}\ee
where the upper limits of $a_2$, $|b_3|$ and $b_4$ come from the unitarity bound~\cite{Lewis:2017dme,No:2018fev}, while the lower limit of $a_2$ is required by a bounded below potential~\cite{Lewis:2017dme}. Note that \Eq{a1} and \Eq{a2} only guarantee $(v,0)$ is a local minimum, and there might be another deeper minimum. For a given set of \Eq{app_input}, one needs to check whether $(v,0)$ is the vacuum (i.e. global minimum) by hand. We found that $\sim38\%$ of the sampling points yield $(v,0)$ as the vacuum. Such points are then phenomenologically allowed.

For the second step, we shift $s\to s+\sigma$ to get the following redefinitions of the coefficients in \Eq{V}~\cite{Espinosa:2011ax},
\be\begin{split}
\mu^2\to&~\mu^2-\frac12a_2\sigma^2-\frac12a_1\sigma,\\
a_1\to&~a_1+2a_2\sigma,\\
b_1\to&~b_1+b_4\sigma^3+b_3\sigma^2+b_2\sigma,\\
b_2\to&~b_2+3b_4\sigma^2+2b_3\sigma,\\
b_3\to&~b_3+3b_4\sigma,
\end{split}\ee
and choose $\sigma$ so that the new $b_1=0$. As a result, the shifted $s$ has a VEV $v_s=-\sigma$. If a negative $v_s$ is obtained, a $\Z_2$ transformation $s\to-s$, $a_1\to -a_1$ and $b_3\to-b_3$ is further performed to make sure $v_s\to-v_s$ is positive. Now the new coefficients combined $v_s$ can be matched to the input parameters in \Eq{input} for the calculation of FOEWPT and GWs. To check the consistency of our treatment, we have verified that the parameters after shifting satisfy the constraints in \Eq{e1} and \Eq{e2}.

\bibliographystyle{JHEP}
\bibliography{references.bib}
\end{document}